# Recovering WPA-3 Network Password by Bypassing the Simultaneous Authentication of Equals Handshake using Social Engineering Captive Portal

**Kyle Chadee**
Department of Computing and Information Technology, St Augustine, University of the West Indies
Email: kyle.chadee@my.uwi.edu
**Wayne Goodridge**
Department of Computing and Information Technology, St Augustine, University of the West Indies
Email: wayne.goodridge@sta.uwi.edu
**Koffka Khan**
Department of Computing and Information Technology, St Augustine, University of the West Indies
Email: koffka.khan@gmail.com

------------------------------------------------------------------ABSTRACT------------------------------------------------------------------

**Wi-Fi Protected Access 3 (WPA3) is the accepted standard for next generation wireless security. WPA3 comes with exciting new features that allows for increased security of Wi-Fi networks. One such feature is the Simultaneous Authentication of Equals (SAE) which is a protocol whereby passphrases are hashed using a Password Authenticated Key Exchange with keys from both the Access Point and the Client making the password resistant to offline dictionary attacks. (Harkins, Dan. 2019) This objective of this research paper seeks to bypass WPA3-SAE to acquire network password via a Man-in-the-Middle attack and Social Engineering. This method can prove to be useful given that majority of network attacks stem from social engineering. For this research we would be looking at the security of WPA3 personal transition mode and capture the network password via a captive portal. Breaching the WPA3 network can be possible by building on various security flaws that was disclosed on WPA3 in 2021. Due to the discovery of Dragonblood downgrade attacks disclosed in 2019, identified that WPA2/3Handshakes could be acquired. A Man in the Middle attack proposed set up is carried out by using race conditions to deauthentication WPA3 network and then using a Raspberry Pi to spawn a rouge WPA3 network. As such, the handshake acquired can then be utilized as to verify the password that would be entered in the captive portal of the rouge WPA3 network. This research identified that the Password was able to be recovered from Social Engineering Captive Portal when Protected Management Frames are not implemented. This research also indicates that some devices are not able to connect to a WPA 3 transition network which is contradicting the Wi-Fi Alliance claim that it is backwards compatible with WPA2.**

Keywords – **WPA3; SAE, Social Engineering, Raspberry Pi, Airgeddon, Captive Portal, WPA3SAE, WPA2-PSK.**

## I. INTRODUCTION

Wi-Fi networks are utilized in almost every industry across the globe. They are utilized to connect everything from databases and Surveillance cameras to Internet of things (IoT) devices and hotspots. Many business functions can be accessed by breaching the Wi-Fi networks. IoT device settings can be altered since they operate on minimal security.

With over 14 years since the introduction of WPA2 being used as the most secure platform for wireless technology (Khan et al., 2023), comes WPA3. WPA3 (Halbouni et. al., 2023) is considered and officially declared as next-gen wireless protocol that addresses the faults of WPA2 and has more secure features. WPA3's incredible features of forward secrecy providing encrypted communication and the Simultaneous Authentication of Equals protocol providing a method of secure connection without transmitting passwords, is hands down a significant achievement for the wireless security community. This research proposes a novel approach to gain the password of a WPA3 network. It is achieved by building on the work produced by many security researchers in order to acquire the WPA3 password in a more effective manner. Many successful computer security attacks are performed via social engineering, and as such the password is acquired through an evil twin spawning a rouge WPA3 network. Forward secrecy protects network traffic from being eavesdropped on and snooped in between the access point and another device.

Even though WPA3 seeks to provide forward secrecy to encrypting traffic, the protocol is still in transition. It would be a while for the full implementation for WPA3 worldwide. However with the evolution speed of technology by the time WPA3 is implemented worldwide, new protocols can be introduced. This research seeks to provide insight into the security of WPA2/WPA3 certification by identifying a mode of breaching the WPA3 network via Social Engineering means. This research methodology outlines the process by which the WPA2/WPA3 password can be breached. It builds on previous researches that identified that the WPA handshake can be recovered from a downgrade attack and then by using a Rouge Access point and captive portal spawned by the Airgeddon Script, recover the network Password. This method is only possible if the Protected Management Frames are disabled in the WPA2/WPA3 transition network. This research seeks to also answer the following questions:

- Can WPA3 password be recovered through a captive portal?
- How long can an attack take?
- How effective would this attack be in a real world scenario?

This research will provide insight into the security of next generation Wi-Fi where it is used from surveillance cameras, operating datacenters, handling Internet of Things (IOT) devices and much more.

The Goal of this research is to breach a WPA3 network utilizing Social Engineering via a Captive Portal and answer the following questions below when incorporated with various ideas to create social engineering attacking platform for WPA3.

- Can WPA3 password be recovered through a captive portal?
- How long can an attack take?
- How effective would this attack be in a real world scenario?
- Developing this method can be able to indicate a vulnerability to WPA3 that may be crucial.

This research aims to investigate the breaching a WPA3 network via a Social Engineering tool, a Captive Portal, spawned on a rouge Access Point. This can provide useful information and control by accessing IoT devices as well as other devices that use WPA2 since WPA3 is susceptible to downgrade attacks. Downgrade attacks can be alleviated but is allowed since it takes advantage of the transition between WPA3 network and WPA2 supported devices. The man in the middle attack that is sought out by this research paper seeks to implement a Captive Portal to trick users into entering Wi-Fi passwords. This work is limited to low cost hardware such as Raspberry Pi's in order to simulate a WPA3 environment. An analysis of this method is performed in order to determine the capabilities and scope of this type of attack against WPA3.

This research hypothesizes that WPA3 transition networks can be breached and have their passwords recovered by building upon and implementing recently disclosed vulnerabilities along with a Captive Portal. This is method is based on Social Engineering attacks performed on WPA2, that can affect WPA3 if utilized together with other recently disclosed vulnerabilities such as the WPA3 Downgrade attack.

A Raspberry Pi were used to simulate WPA3 environment using OpenWrt. This is achieved by a three step process where the real WPA3 networks is downgraded to WPA2 and the handshake is acquired then a Rouge WPA3 network with a captive portal is spawned to verify the handshake acquired to what the user would enter in. In analyzing this method it may be more effective than using brute force means to recover the network password under certain circumstances.

The research is conducted under the following Assumptions and Limitations:

- It is assumed that Protected Management Frames would be off in the WPA3 settings.
- This is plausible since users may not be cognitive of the settings of WPA3.

Due to the strength of WPA3 thus far, Security researchers are working assiduously around the world to enhance and fix all flaws in WPA3. The proposed Denial of Service operates only if certain assumptions are made. These assumptions are: The protected frame management is disable and that access point CPU cannot handle more than 16 frames/ bad tokens.

The Deauthentication attack allows for users to be disconnected so that they can renter the password within the captive portal. However, other approaches to deauthentication still prove useful to achieve the Denial of Service but the attack can be performed without it preying on newly connected devices that don't auto connect to the network.

Cost of devices to be purchased to spawn rouge networks and acquire password is not much and as such devices such as Raspberry Pi's and Wi-Fi pineapple can be configured to run the WPA supplicant but it's still a cost factor.

This research contributes to research on security of WPA3, and more security concerning specifically recovering the network password. It shows that the network password can be recovered from using a Social Engineering Captive Portal. It also shows other important discoveries where Only laptops that are compatible to connect to the transition network can be utilized even though research articles and the Wi-Fi Alliance stated that the transition network, WPA2/WPA3, is backwards compatible to work with WPA2. It is also discovered that two messages out of four messages in the WPA handshake can be recovered if a WPA3 client device connects to a WPA2 Network as oppose to the reverse as outlined in the recently discovered flaws of WPA3.

## II. METHODOLOGY

As stated before, WPA3 is the frontier for next generation wireless security (Sun, 2021). However, when talking about technologies, security is derived from how long a technology takes to be breached. In attempts to achieve strong security, constant testing methods have to be conducted to ensure that minimal, non-critical loop holes exists. To enhance the security of WPA2/WPA3 certification, this project seeks to identify a mode of breaching the WPA3 network via Social Engineering means.

Penetration Testing allows for vulnerabilities and zero day faults to be detected. Breaching the WPA3 network can give insight into the weaknesses of the network and as such, patches or various solutions can be derived in order to alleviate the weakness and make the system much more secure. A breached Wi-Fi network can be a cause for concern. Many criminal activities can take place if a threat actor takes control over the Wi-Fi network. Some of these activities can even be very drastic and can reach up the ladder to Fraud and even terrorism. By being able to breach the network, access to various IoT devices and personal computers can become easier since scanning the network to see what vulnerable devices are on it.

This research paper seeks to also build upon the research conducted by other security researchers on the security of WPA3. This research seeks to also answer the following questions:

- Can WPA3 password be recovered through a captive portal?
- How long can an attack take?
- How effective would this attack be in a real world scenario?
- It also seeks to incorporate these various ideas to create social engineering attacking platform for WPA3.

Breaching the WPA3 network cannot be easily done due to its increased security features and due to little research on WPA3 exists, in comparison to WPA2, since its next generation Wi-Fi security. One method to breach the WPA3 network is by the taking advantage of attacks and vulnerabilities previously discovered by Mathy Vanhoef on WPA3. One method is the use a downgrade attack to recover the handshake and use brute force to crack it. Another method is to use of Social Engineering and utilizing a Captive portal. Here targets can enter their password and have it verified. This research paper seeks to also build upon the research conducted by other security researchers on the security of WPA3. After the WPA3 handshake is recovered, three possible options can be used to deauthenticate legitimate clients from the real access point and entice connection to the captive portal. (Dabrowski et al., 2016) These three methods to create a deauthentication attack can be to use a Race Condition, Overload the Access point with requests or to utilize a Bad token.

With this in mind, using a downgrade attack to capture the WPA handshake and then use a Bad Token Deauthentication attack to jam the WPA3 network (providing a Denial-of-service) and then spawning a captive portal to capture the Password and verify it against the captured handshake. This method is what selected to breach the WPA3 network that would aid in strengthening the security of the WPA3 certification.

This attack is based on various research work performed and then by building on those research works, develop a strategy to acquire the WPA3 handshake. Mathy Vanhoef and Eyal Ronen proved that WPA3 handshake can be downgraded to WPA2 and captured as well as Karim Lounis and Mohammad Zulkernine proved that through Race Condition and Bad Tokens a Denial of Service Attack can be performed on the WPA3 network. Due to the strategies utilized by Fluxion and Airgeddon, where the handshake is acquired and then the legitimate devices are deauthenticated and denied reconnection from the real access point in order to connect to a rouge access point. This access point will have the same SSID that would have a captive portal to enter their Wi-Fi password. From there the captured handshake from before would be used to verify the entered password and therefore determine if it's valid or not. A similar approach can be used to acquire a WPA3 Password from a Captive Portal if these research articles were built upon together. The attack is performed in three steps (see Figure 1).

1. Handshake capturing using downgrade attack
2. Deauthentication Of Original WPA3 network and then
3. Spawning an evil twin and Captive portal to get the password.

Research performed by Vanhoef et.al, indicated that a downgrade attack on WPA2/3 transition networks resulted in the handshake being captured. This is due to backwards compatibility not being able to be achieved in WPA3 and as such if a supplicant that operates on WPA2 tries to connect to WPA3 the handshake would be revealed. The captured handshake is then utilized to verify the authenticity of the password that would be entered in the captive portal in the third step.

After capturing the three way handshake, Deauthentication attacks will be performed against the original WPA3 network. Since WPA3 is highly secure due to Simultaneous Authentication of Equals and Protected Management Frames, Deauthentication attacks can be trivial to accomplish. However, both researches conducted by Vanhoef et.al 2021 and Lounis et.al 2019 , we see that there are three methods by which a deauthentication attacks can be possible if protected management frames are not implemented within the original WPA3 network. Vanhoef and Ronen indicated that due to CPU performance of the Access Point, requests can be sent to overload the Access Point and thereby providing a Denial of Service attack. He also suggested another option to deny legitimate clients from connecting to the WPA3 access point where a race condition between a WPA2 and WPA3 network can take place if a WPA3 device connects to a WPA2 access point. (Vanhoef and Ronen, 2021)

Lounis et.al 2019, indicated that Bad Tokens can be propagated to create a Denial of Service, by utilizing this method a Denial of Service can be executed upon the original access point thereby allowing room for a rouge WPA3 network with a captive portal to be created and easily connected to since users would not be able to connect to the legitimate access point. The authors also indicated in their research on WPA3 connection deprivation attacks 2020, that there are three possible ways that can be used in the deprivation of authorized users from connecting to the WPA3 network. The user will then enter the password which would be then verified with the handshake acquired in the first step.

## System

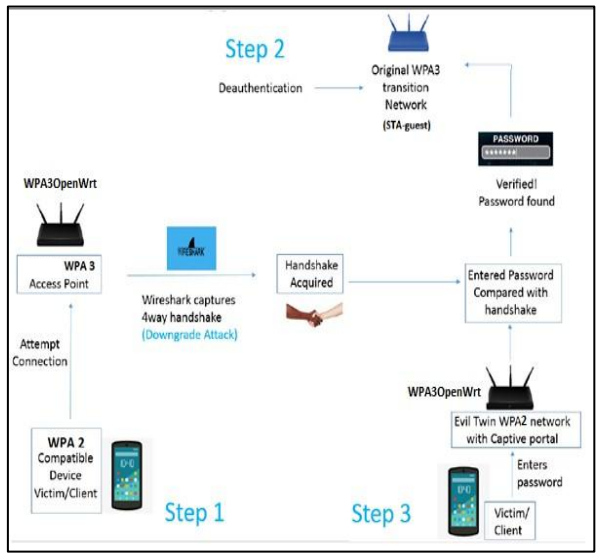

Figure 1: Showing how the proposed research idea would be set up to acquire the WPA3 password

This method was selected for various reasons. The first is that the Bad Token research conducted proved useful results against a majority of WPA3 access points. The basis of the attack was substantial in their research paper and the fact that many people cited this article can prove and attest to its creditability.

Attacking the access point to overload it depends on the processing power of the CPU. Since it's considered cheap to get cloud resources online to generate enough requests to overload the Access point, makes it a viable option to be utilized as the deauthentication mechanism to provide a Denial of service attack.

The WPA3 handshake contains 2 out of the four frames used in the four-way handshake of WAP2 by using a downgrade attack on a WPA2/WPA3 transition network, recovering two out the four frames is possible and thus it can be brute forced or utilized in a captive portal. A captive portal proved much more effective than using brute force. This is the main advantage of using the Captive portal in order to cut down the time taken to acquire the password from days to minutes.

The work produced by Mathy Vanhoef is critical but the effectiveness in recovering the password via brute force can take days depending on size of the dictionary or word lists used, that's if the password is even listed in there in the first place. With this method of using a captive portal, a different approach to capturing the accurate password can be performed, and depending on the situation environment, it can even be a faster to acquire than using brute force (Bhatt, 2021). However, it must be noted that using brute force attacks are completely silent in terms of offline dictionary attacks. When it is compared to using a captive portal, the captive portal involves user interaction thus it is technically revealing the attack. People who are educated on cyber security can pick up on the captive portal.

To answer the questions set out in this research, an experiment or proof of concept has been set up. Some hardware equipment that would be necessary in this experiment would be a Raspberry Pi, a Laptop with Kali Linux installed and two mobile devices such as two cellphones. Some software equipment that would be necessary is Wireshark, Aireplay, Airgeddon and OpenWrt.

This experiment would feature a Raspberry Pi to simulate a WPA2/WPA3 transition network which would be attacked. Using a laptop, Wireshark is utilized to capture packets and show the live capturing of the WPA handshake after downgraded. Aireplay is used to capture the WPA handshake which is then stored in a system folder. The captive portal within Airgeddon is then utilized to simulate the Rouge Access Point with a captive portal where unsuspecting users would connect to and enter their password. Capturing the Handshake is performed by using the downgrade attack on WPA3 mentioned by Mathy Vanhoef. He stated that a WPA3 device connecting to a WPA2 Access point would cause the Access Point to revert to WPA2 functionality. This research seeks to do the reverse and use a WPA2 device connecting to a WPA3 Access Point to theoretically do the same. This is investigated in this research as well.

The use of Aireplay tool in Kali Linux will be used to capture the handshake. Airgeddon contains a suite of network auditing strategies. It can create deauthentication attacks against WPA2, capture WPA2 handshakes and spawn rouge Access Points with sniffing enabled. For this research, Airgeddon Evil Twin access point with captive portal will be used. The captive portal would be a browser page that has an input box instructing the unsuspecting victim to input the network password. This tool also verifies the password with the handshake acquired to ensure that the network password is correct. (da Silva, 2023)

The Denial of Service attack can be performed using the strategies such as Bad Tokens, to create a Denial of Service on WPA3 as mentioned in the journal Bad Token: Denial of Service attacks on WPA3 by Karim Lounis and Mohammed Zulkerine, 2019 or the "DragonDrain" exploit derived by Mathy Vanhoef and Eyal Ronen, 2019.

The Denial of Service attack outlined by Mathy Vanhoef and Eyal Ronen is in the form of a Resource consumption attack. This stems from the CPU usage of the Access point. When the commit frame is being processed by the access point, generating a hash can be computational expensive that would result in an overload. This is due to the cookie exchange method in WPA3 that prevents devices from using fake MAC addresses to forged commit frames to generate a computational expensive answer. Because of this, the Access Point can be overloaded by a minimum of 16 forged commit frames utilized per second. This can freeze the Access point or slow it down significantly, either way is causes a denial of service to devices. Dragon Drain is a script composed by Mathy Vanhoef to exploit this vulnerability.

The other Denial of Service attack on WPA3 that Karim Lounis and Mohammed Zulkerine suggested is to launch an attacker supplicant when a legitimate supplicant is about to connect. The two devices would send commit frames to the Access Point which would then reply with a confirm message. The two devices would then send their confirm messages in which the Access Point would receive the first one from the attacking device and send an error code 0x0001 which is "Unspecified Failure" to the legitimate device trying to connect. This connection would then be aborted. The authors of this research paper claimed to have this attack repeated in a synchronous way for one hour thereby creating a Denial of Service attack for an hour against the WPA3 device.

For this research paper, the first option provided by Mathy Vanhoef and Eyal Ronen is examined to provide the Denial of Service in this attack. This is due to the likelihood of resource consumption attack appearing more successful since the access point used in this experiment is a Raspberry Pi. An article posted on July 6th, 2021 by Fusion Connect, on their website indicates that Normal Access Points can typically handle 45 connections even though the optimal claim might be around 250 devices. However with the Raspberry Pi, it has a 32 GB memory card. This can affect the number of devices connected to it. As such, it seems more plausible that the Resource Consumption attack might be more successful for this experiment.

Design goals included the following:

• Cost: In conducting this experiment, it is tailored to such a way that it will operate with a minimal budget. This means that instead of purchasing WPA3 routers, Raspberry Pi's were used to simulate WPA2/WPA3 network.

• Reliability: In order to ensure that the experiment produce reliable results, the experiment was conducted with multiple cellular devices to ensure handshake capture. Due to the robustness and reliability of OpenWrt, this method was selected to use as the WPA3 access point instead of using Hostapd directly.

• Portability/Location: In a real world scenario, attacks performed against the WPA2/WPA3 network may not be where the attacker and Access Point is situated direct besides each other. As such a distance of about six feet is place between the Access Point and the attacking laptop as a means of simulating an overhead WPA2/WPA3 Wi-Fi router.

The equipment sought for this experiment was sourced on Amazon.com and Ebay.com on 17th November 2021.

Table 4: Showing the Cost of the Equipment utilized in this research

| Equipment | Model and Specifications | Cost (USD) |
|---|---|---|
| Laptop | Lenovo ThinkPad I7 | $600.00 |
| | HP Pavilion I7 | $669.00 |
| Raspberry Pi 3 | Model B+ | $89.00 |
| Cellphone/Tablet | Samsung J5 2016 (Android 7.1.1) | $70.00 |
| | Huawei Ascend Mate 7 (Ebay) | $109.00 |

We will not discuss the software design. OpenWrt image was installed into an SD card and uploaded into the Raspberry pi. Then using the laptop SSH into the Raspberry

Pi and create and WPA3 Access Point.

OpenWrt user interface made altering configuration easy and changing the password and settings.

Airgeddon is a wireless auditing script that would allow a Rouge Access Point and captive portal to be spawned to acquire the WPA2-PSK/WPA3-SAE password (CyberPunk, 2020).

Since the attack is centered on user interaction, client's interaction may vary. Clients can become very active, mild active or not active in interacting with the Rogue Access Point. This alters the time frame in determining how successfully the attack would be.

It is expected that using the captive portal would reveal the WPA3 password by matching the captured handshake against the entered handshake.

Due to research conducted by Mathy Vanhoef, one of his experiments on the Downgrade attacks utilized a WPA3 device trying to connect to a WPA2 Access Point where the handshake was captured. In light of this, it was therefore assumed that the reverse can be performed and still capture the handshake since Wireshark in monitor mode would pick up transmissions over the air. It is assumed that since Mathy Vanhoef was able to acquire the handshake from a client device with WPA3 and Access Point with WPA2, the reverse where a client WPA2 device trying to connect to a WPA2/WPA3 transition network work also work. (Goyal and Goyal, 2017)

In installing the OpenWrt firmware on the Raspberry Pi, problems in the connection can occur. The connection can drop or become disconnected when using SSH to access the firmware. This can be due to the power input of the Raspberry Pi not supplying the required voltage that wouls be necessary to keep the connection properly working. Reducing the amount of devices connected to the Raspberry Pi WPA3OpenWrt access point can help reduce this concern.

### III. RESULTS

WPA3- certification provide additional security to networks and has been rolled out since 2018. At this point there are many WPA3 access points being used around the world as well as WPA2/3 transition access points. This experiment acts as "Proof of Concept" to the strategy outline in this paper. The aim of this experiment seeks to breach a WPA2/3 transition network and recover the networks password with the use of a rouge access point with captive portal. This experiment is performed in three stages:

• Handshake capturing using downgrade attack
• Deauthentication Of Original WPA3 network and then
• Spawning an evil twin and Captive portal to get the password.

In order to perform this experiment a list of materials is first needed. These items were listed as both hardware and software requirements in the previous section, Methodology. These are:

• 2 Raspberry Pi's
• Monitor
• Keyboard and Mouse
• 2 Memory cards at least 64 GB (microSD)

• Laptop with Kali Linux and at least two Wi-Fi interfaces/dongles
• Airgeddon
• Aireplay
• Wireshark
• Internet connection
• 2 Cellular phones
• Raspberry Pi Writer tool
• OpenWrt

We now give the real-world implementation with details.

*1) Setting up the raspberry Pi with OpenWrt*

The first stage of setting up the system is to install the OpenWrt. OpenWrt firmware acts an open source custom firmware for routers, however for this experiment an OpenWrt image is installed on a raspberry Pi for it to act like a WPA2/3 transition router. This is achieved by first downloading the OpenWrt latest image from their website. Then utilizing the raspberry Pi writer tool, the image was written to a Micro SD card. This Micro SD card was then ejected and placed into the Raspberry Pi's memory card slot. The peripheral devices such as the mouse, keyboard the monitor were connected and then the power supply was turned on. A Wi-Fi dongle was also attached to give internet connection to the Raspberry Pi. The Raspberry Pi then boots into the OpenWrt firmware and a command line interface is show. Initial settings were then configured.

Using the "Creating a WPA3 Access Point using OpenWrt on Raspberry Pi 3B+" tutorial by Wolker Lemahieu, the WPA3 access point was set up. A RJ45 Ethernet cable was attached from the Pi onto a laptop where it used SSH to access the firmware and configure files. The Static IP of the Pi was then altered from 192.168.1.1 to 192.168.1.2 since the router connected to the Pi through the Wi-Fi Dongle is already in use. This was done by using the built in shell provided by the Raspberry Pi by the following codes.

*a)* *Table 1: Shows below show the commands to change the static IP address of the OpenWrt in the Raspberry Pi*

| uci set |
| --- |
| network.lan.ipaddr=192.168.1.2 |
| uci commit |
| /etc/init.d/network restart |

Adapted From: Lemahieu, Wolker. 2016. " Create a WPA3 access point using OpenWrt on a RASPBERRY PI 3B+". *Wolkerlemahieu.Be*. https://wolkerlemahieu.be/tutorial/tutorial.html.

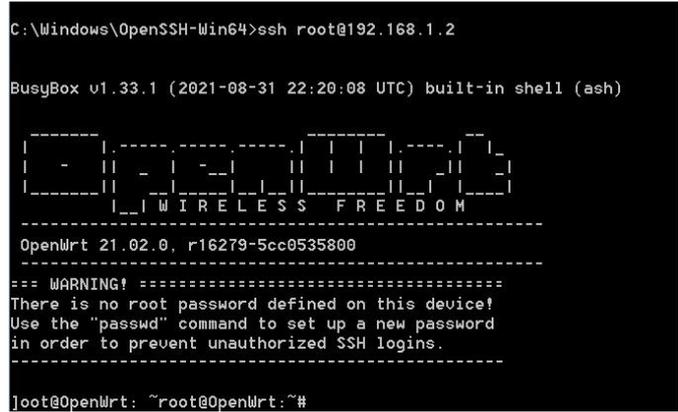

Figure 2: Showing the OpenWrt on the Raspberry Pi is connected to via SSH

*2) Setting up the WPA3 Access Point*

Once raspberry Pi is set up, now it's time to create the access point. The access point was configured using the same "Creating a WPA3 Access Point using OpenWrt on Raspberry Pi 3B+" tutorial by Wolker Lemahieu.

A browser was then opened and the home address of the OpenWrt on the Raspberry Pi, 192,168.1.2, was navigated to. The password was then set for the admin page of the OpenWrt settings. Though this tutorial was followed to set up the WPA3 network, it must be noted that not all the steps mentioned in this article was taken as some steps provided errors in configuration.

The Network tab was then selected and the Interfaces section was selected. The Edit option was then selected. In this section, the IPv4 and DNS address was kept the same, even though the Article would have indicted to set them both as the address of the router, 192.168.1.1. This was acceptable since this experiment does not need internet connection, but only to simulate a WPA3 network. However, it must be noted that it can be done even though.

The DHCP tab was then selected, and the ignore interface box was checked. This would disable the DHCP server for the interface as the router would be doing that (Alghisi, 2024). This was only applicable if the Ipv4 and the DNS address were set to 192.18.1.1. IPv6 setting was then chosen on the tabs of the window and all the IPV6 settings were disabled. The save and apply button was then clicked to save the settings.

Some services in OpenWrt can be disabled if not utilizing the internet or if these services are using the settings of the router connected and not those listed in the OpenWrt. For this experiment, Internet was not connected to so some settings were disabled. These were the Firewall, DNSMASQ, and the ODHCPD while the network option was kept enabled. The Raspberry Pi system was then restarted so that changes would take effect.

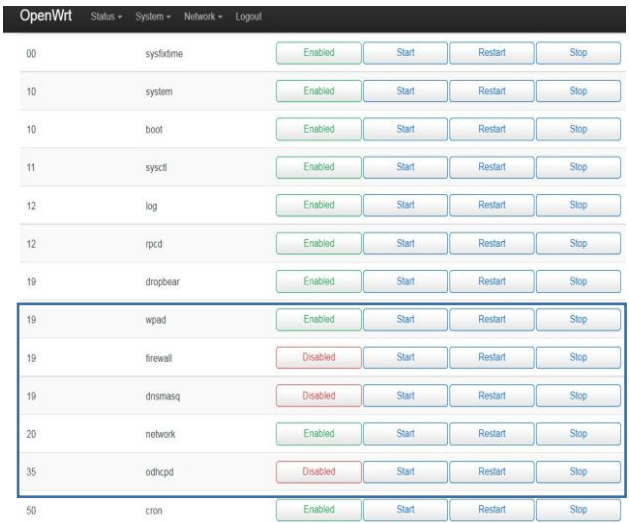

Figure 3: Showing the services being disabled to allow proper functioning of the WPA2/WPA3 Access Point

After the Restart, the Network tab was then selected, the interfaces option was selected and the second option being the second network was then enabled and edited. The SSID was then changed to "WPA3OpenWrt."

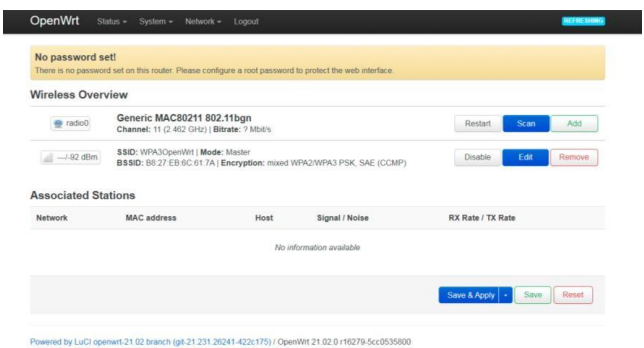

Figure 4: Showing the wireless interfaces in OpenWrt

The Encryption mode was then set to WPA2-PSK/WPA3-SAE Mixed mode to enable the WPA3 settings and make the router WPA3 compatible with WPA2 backwards compatibility. However, WPA3 enterprise is not backwards compatible so only in the transition mode is where this research seeks to apply this vulnerability. (Kwon and Choi, 2020)

This option was only available in the OpenWrt version 21.02.0 version and above, otherwise the package to incorporate WPA3 would need to be installed. This is done by going to the System tab, clicking software option and updating the Lists. The result option that is labeled "hostapdopenssl" would then be installed and the WPA3 settings would be present.

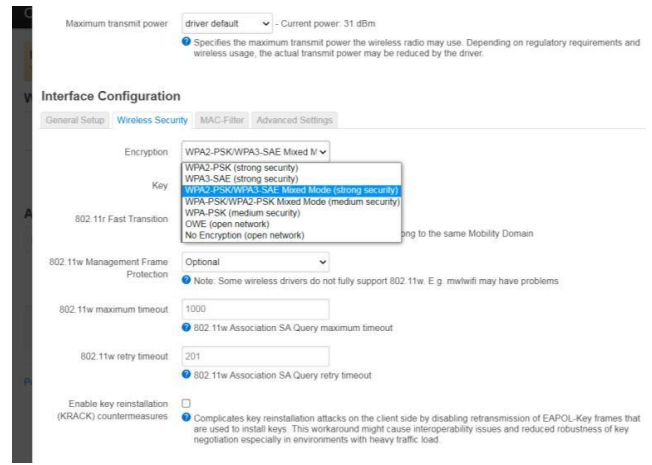

Figure 5: Showing the WPA2-PSK/WPA3-SAE mixed mode being set in the OpenWrt Access Point

The Password was then set for the WPA2-PSK/WPA3-SAE transition access point. The password set up for this demo was "12345678."

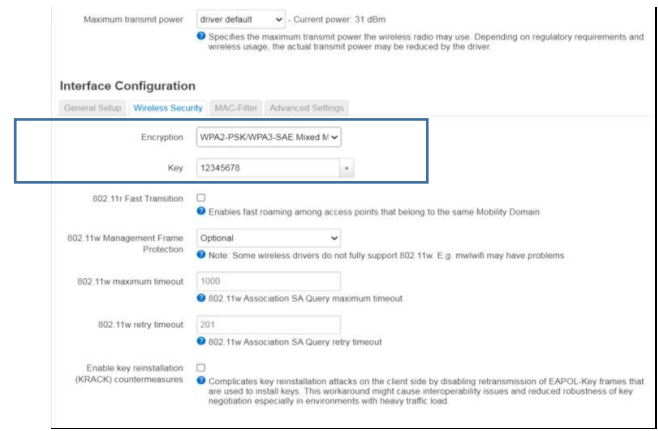

Figure 6: Showing the password being set for the WPA3 access point

When this was finished, the Wireless Access point is now active and is picked up by nearby devices.

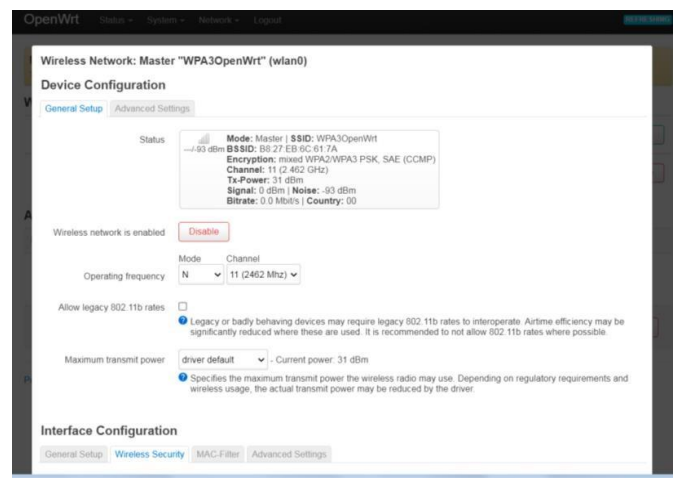

Figure 7: Showing the final settings of the Configure WPA2/WPA3 Access Point

The same procedure is performed to set up a WPA3 device supplicant to connect to the WPA3 network. Wireshark is started in monitor mode to identify that the SAE is in effect and "Aireplay" script was executed to show that the Handshake could not be captured.

### 3) Capturing the Handshake

The first step in capturing the handshake is the Deauthentication attack. Three strategies were outlined earlier on how this can be achieved supported but literature. The deauthentication attack is first performed on the WPA2/3 access point. Seeing that the WPA2/3 transition access point also incorporates the functions of WPA3, the denial of service attacks listed against WPA3 would work here as well. This would cause both WPA2 users and WPA3 users to loose service thereby leading the uses to think it's a network connection problem and not an attack to deny them of service. This would then cause the WPA2 users to try to reconnect.

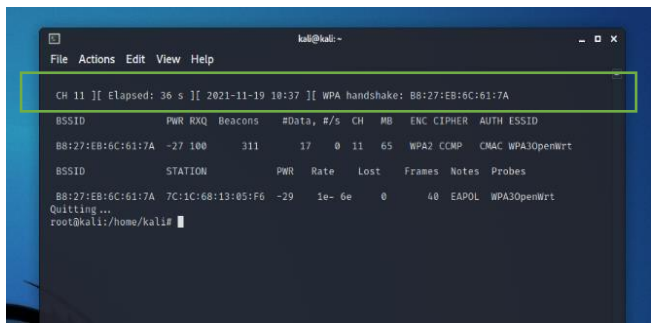

Figure 8: Showing the Downgraded WPA3 handshake being capture by Aireplay

For this section of the demonstration, a cellular device that is WPA2 compatible is used to connect to the WPA2 access point. Due to the downgrade attack being discovered, the WPA2/3 Access point would downgrade its WPA3 settings to accommodate the WPA2 device. Aireplay and Wireshark has started to show that the handshake has been downgraded to two messages and is captured.

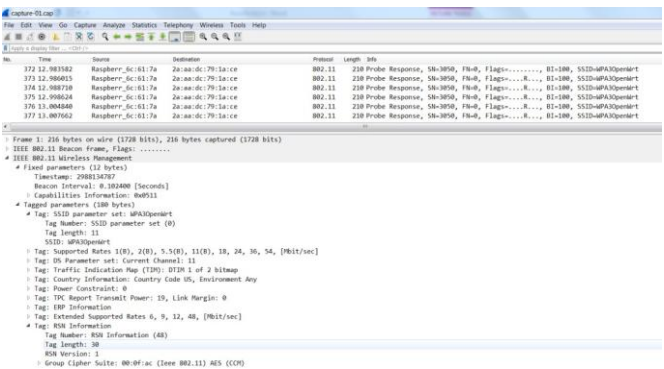

Figure 9: Showing the monitoring of the WPA3 network.

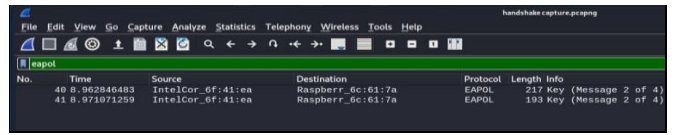

Figure 10: Showing 2 out of the 4 handshake messages being captured in Wireshark Monitor mode

### 4) Setting up the Rouge Access Point

The Airgeddon program is downloaded from GitHub and installed on the Kali Linux distro. The program then selects one of the connected wireless interfaces to operate as an Access point. Then using Hostapd user space daemon utilize the network interface card and enable it to work as an access point. The Airgeddon program then takes the captured handshake as a parameter, and spawns the captive portal based on that for unsuspecting users connect to. When they connect to it, they type in their password and it is verified by using the captured hash. This is how the WPA2-PSK/WPA3-SAE Network Password can be recovered. (Liu,2019)

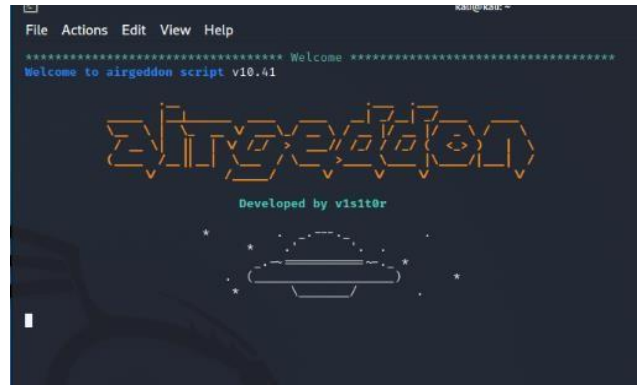

Figure 11: Showing the Airgeddon script and the version being used

In setting up the rouge access point to gain the password for the WPA2/WPA3 access point, the script Airgeddon is used. This program would generate a WPA2 Rouge Access Point with a captive portal interface. Since the security of WPA3 is enforced on the Access Point, the WPA3 device will still connect to a WPA2 Access Point.

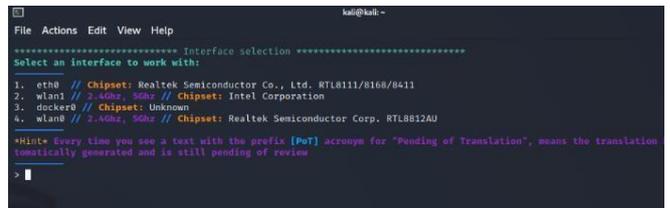

Figure 12: Showing the wireless interfaces than can be used by the Airgeddon script

The first setting in the Airgeddon is to select an interface and place it into monitor mode. For this research the interface wlan0 is used. The Realtek Semiconductor Corp. Antenna RTL8812AU is utilized.

Figure 13: Showing the Interface selected and setting it into monitor mode

Figure 14: Showing the Evil Twin Access Point with Captive portal option being selected

After monitor mode is initiated, the option to generate the captive portal was selected at option 9. Using the interface that was in monitor mode, the surrounding area was scanned for nearby networks, and as such the WPA3OpenWrt network was identified. This network was then selected, as option 80, to spawn an attack against. In the image below, the other networks and their relevant information was hidden for privacy reasons.

Figure 15: Showing the WPA3OpenNetwork being listed in the network scan as the network to be attacked as option 80

The Red blocks seen in Figure 15 are there to block out other SSID's and MAC addresses in an attempt to maintain privacy and security of those networks since they are not being utilized in this research.

Figure 16: Showing the MAC address spoofing by Airgeddon

In selecting the WPA3OpenWrt network, the script can engage in MAC spoofing of the legitimate access point WPA3Openwrt. The option for MAC address spoofing was initiated and the path to the handshake that was acquired earlier was entered as well. The MAC spoofing allows for other devices to see the rouge access point with the same MAC address as the legitimate access point. (Benzaid et al., 2016)

Figure 17: Showing the WPA3 captured handshake being used in the attack

After the path to the capture WPA2/WPA3 handshake was entered, the path to store the capture password is entered as well. This is stored on the desktop as a text file named WPA3OpenWrtPassword.txt.

Figure 18: Showing the menu option to select the language of the Captive Portal to use

The Airgeddon menu options has a variety of languages by which a captive portal can be spawned. The English language was selected and utilized to spawn the captive portal.

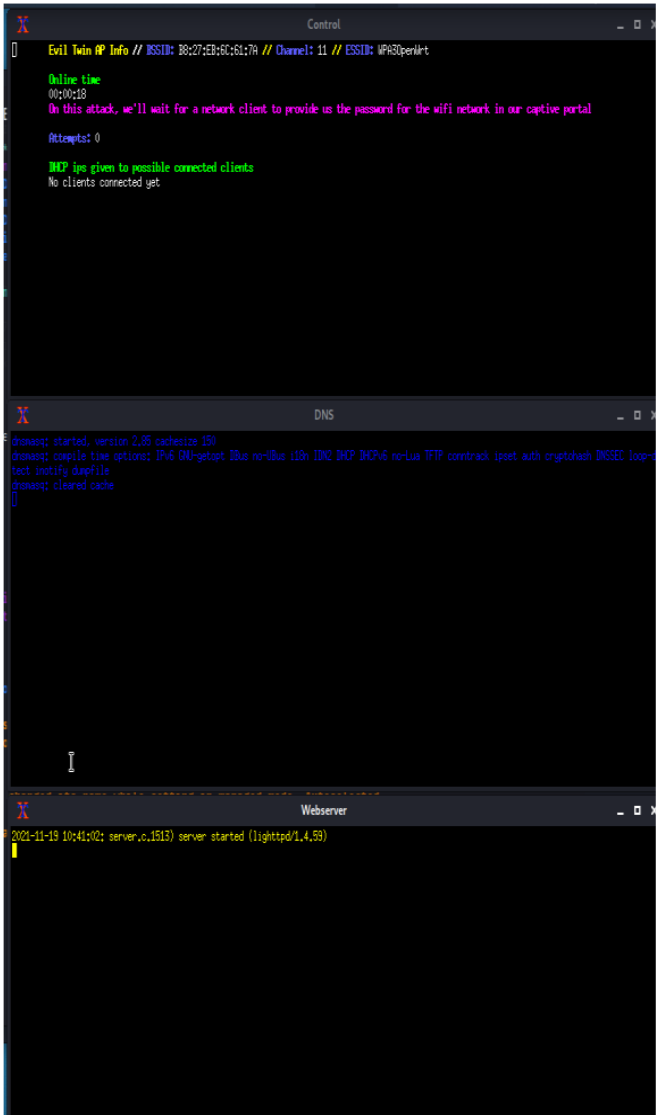

Figure 19: Showing the Rouge Access Point being started with the Captive Portal from the Attacker Side

Once the Rouge Access Point with the Captive Portal was spawned, six frames are spawned. The access point frame that monitors connection to the Access point, The DHCP frame for the Access Point, A control frame that monitors to see if the password was entered, a DNS frame for the Access Point, and a webserver. The deauthentication frame is also initiated, however, WPA3 is susceptible to the wpa2 Aireplay deauthentication attacks. So this frame blocks it. However in a research performed by Dalal et,al. On Wireless Intrusion Detection system for 802.11 WPA3 Networks in October 2021 from the Cornell University in the USA, indicated that a deauthentication attack that is possible in WPA2 networks is still possible in WPA3 therefore meaning that the WPA2/WPA3 network was susceptible to deauthentication attacks.

At this point, a WPA3 deauthentication attack can be used. This can either be using the Bad Tokens to create a

Denial of Service on WPA3 as mentioned in the journal Bad Token: Denial of Service attacks on WPA3 by Karim Lounis and Mohammed Zulkerine, 2019 or the "DragonDrain" exploit derived by Mathy Vanhoef and Eyal Ronen, 2019. These exploits are published and credible modes of deauthentication and are recognized internationally. However due to Time constraints and configuration of these tools, the deauthentication exploit was not created. It must note that the deauthentication attack was only to cause WPA3 devices to be inclined to join the Rouge Access Point when they can't get access to the legitimate Access Point, but it is not necessary to do so.

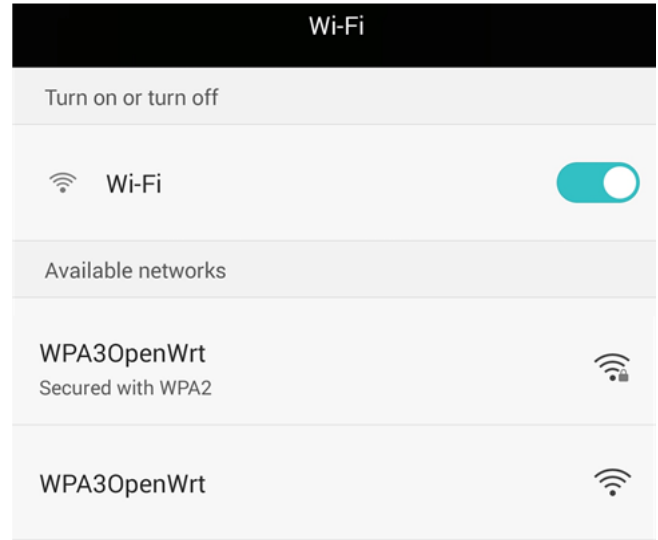

Figure 20: Showing the Original WPA3 transition network and a Rouge WPA3 transition network with the same SSID operating at the same time

Once the Rouge Access Point was spawned, both the rouge Access point and the legitimated ones showed up in the Wi-Fi scan. The image above clearly shows that the Rouge Access point utilizes WPA2 security. This however, is only identifiable in some devices as some devices would not even show that distinction.

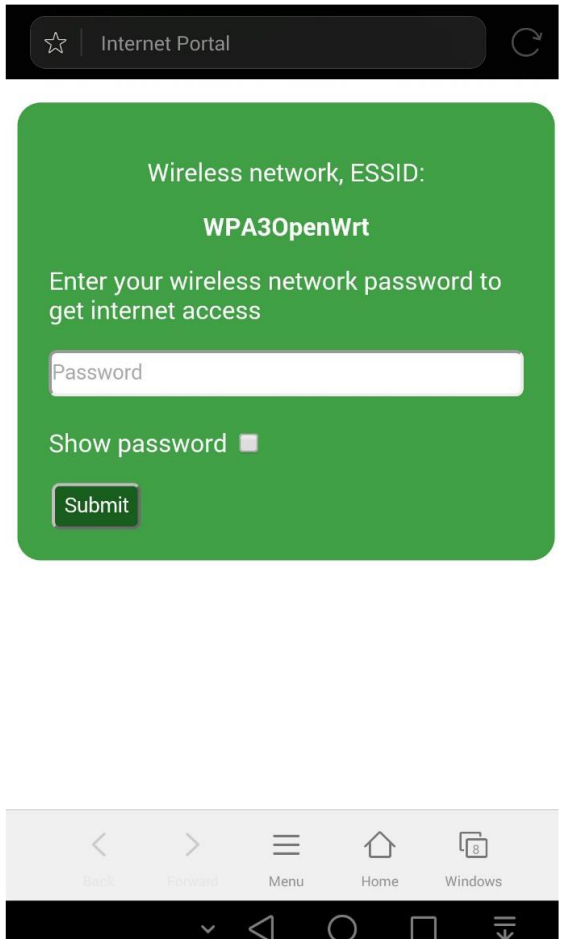

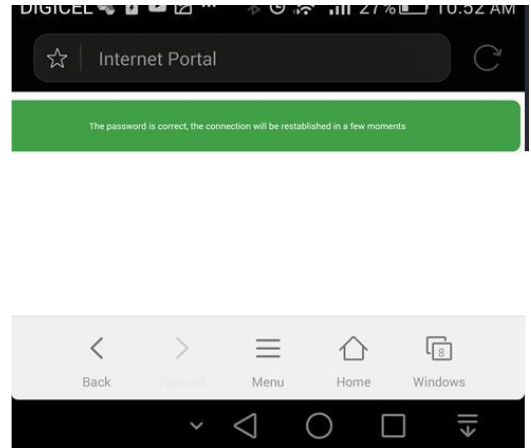

Figure 22: Showing the Fake Connection message on the Client celular device

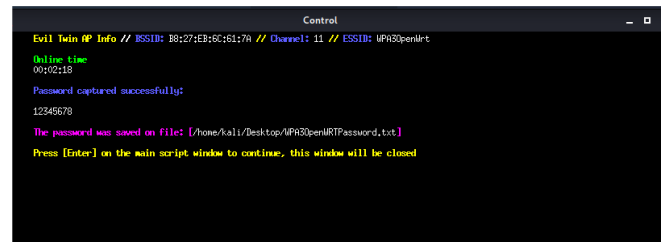

Figure 23: Showing the WPA3 transition network Password being captured and logged

## IV. DISCUSSION

This research sought to recover the WPA2/WPA3 password by the use of a Captive Portal. The objective of the experiment was to recover the WPA3 network password which was identified from the experiment. Some open source solutions can be used to simulate the WPA3 access point such as OpenWrt and Hostapd (Hätönen et.,al, 2016). Using OpenWrt instead of installing Hostapd directly on the Raspberry Pi provided the added advantage of having a GUI browser interface. This interface made it easier to navigate through and set up the WPA2-PSK/WPA3SAE network. Writing the OpenWrt Image to the SD card produced was repeated a number of times due to the SD card not being able to read properly. It is noted that the quality of the SD card affected the desired result of placing operating images onto the Raspberry Pi. This can be a serious issue in the execution of the system since it's not like having a PC or other device where memory and CPU's can be added on.

The memory size of the SD card used in this experiment was 32 GB which is enough to house the OpenWrt image that was around 180,000KB. This would allow a substantial amount of devices to connect to the rouge access point in order to retrieve the password via a captive portal. It must be noted that the use of the Raspberry Pi in this experiment was for demonstration purposes, but if using a larger size SD card with good quality, many as 5 devices can connect to the Raspberry pi when acting as an access point. Due to its size and portability makes it the excellent choice as well for on the field activities and can be used in real world attacks.

Raspberry Pi's utilize a voltage of around 5 Volts plus or minus 5% in order to perform its functions. When the

Figure 21: Showing the Captive Portal page of the Rouge Access Point on a Cellular Device where the client would enter the WPA3 Password

Once a User accesses the Rouge Access point, the device would be redirected to the web page containing the captive portal. The Captive Portal can be configured to resemble of look like any router home page but for the sake of the experiment it was kept simple. When the unsuspecting user enters the password in the captive portal as shown in the Figure above, the password is then revealed. The captive portal can then alerts the user with a fake message indicating that the connection was established to the network and then the attack is closed off. This means that the captive portal and the denial of service attacks against the legitimate WPA3 network is closed off. The revealed password is also logged in the WPA3OpenWrtPassword.txt file on the Desktop.

Raspberry Pi operating system was Pi was connected to a power source, the Pi indicated a low voltage reading. The Memory card of the Raspberry Pi was then wiped from it Raspberry Pi operating System and OpenWrt was uploaded onto it to create the WPA3 access point. According to Vladimir Vujovic and Mirjana Maksimovic, in their research article on Raspberry Pi as a Wireless Sensor, 2014, indicated that the increasing the rate of communication when using a Raspberry Pi has significant impact on the power. In light of this, care was taken to ensure that the Pi is not overloaded by having too many devices connected to it in this testing phase. (Thurium, 2019)

In capturing the WPA2/WPA3 handshake, unsuspecting users might become suspicious of it. One way of reducing the doubts that the client may have in connecting to the network due to this is to alter the Captive portal page to make it look almost exactly like the brand/home page of the router. (Jain et al., 2019) Having the captive portal resemble the original network can increase the chances of the user entering the password. For example. The original network device brand that a company is using might be a Cisco Access Point in English language. Formulating the captive portal to appear that it's a Cisco device and the language in English might be more successful than spawning a captive portal to appear like a Huawei network in Russian language. This means that the effectiveness of the captive portal can thus be affected by the design of the Captive Portal. However, to figure out the brand name of the Access Point being used by the victim, a Wireshark scan in monitor mode of the surrounding area can reveal the brand name of the device being used. This information can then be used to construct the captive portal to increase chances of success. During the experiment the captive portal was able to determine the WPA3 password. The Captive Portal was able to match the entered password with the captured handshake thereby confirming that the password is authentic.

During the capture of the WPA2/WPA3 handshake via a downgrade attack, it was noted that EAPOL messages captured by Wireshark can be malformed. This was identified when the HP pavilion Laptop tried to connect to the WPA2-PSK/WPA3-SAE network.

Reaction to the captive portal varied from device to device. A Huawei Ascend mate 7 cellular device was used. The cellular device showed that the WPA3 original network has "None" for its authentication type even though the Access Point contained WPA3 authentication. This occurred due to the cellular device not being compatible with WPA3 and as such didn't not know how to classify the authentication method. When the network was observed using the Lenovo Laptop, the WPA3 network was listed as having an encryption type of WPA2. This is due to the backwards compatibility of the WPA2/WPA3 network. However, another laptop was tested out against it as well, a HP pavilion, and it was discovered to no be compatible with the network and not connection was acquired. In a post on the Wi-Fi Alliance's website on June 25th 2018 indicated that WPA2 devices can continue to be utilized and interoperate with the recognized security. This means that WPA2 can still be utilized with the WPA2-PSK/WPA3-SAE transition networks. This shows that not all WPA2 devices are compatible with WPA2/WPA3 transition networks even though research articles and legitimate websites such as the Wi-Fi.org which is owned by the Wi-Fi Alliance, claimed it to be.

The results of this experiment were in support of the expected results outlined by this research. It is evident that a captive portal can in fact be utilized to capture the WPA2/WAP3 network password. It was also evident that using a WPA2 device to access a WPA2-PSK/WPA3-SAE network, a downgrade attack would occur. This was first seen in the research paper produced by Mathy Vanhoef and Eyal Ronen which outline the downgrade attack being done in reverse with a WPA3 device and a WPA2 Access point. This research proved that the reverse can be done as well since the communication between the devices is what is being monitored and weak. The information transmitted was the target and not just the devices themselves. This lead to the assumption that capturing the handshake using the reverse strategy can work as well. The captive portal has success against WPA2 networks and now it has success against WPA2/WPA3 transition networks. It is expected that the WPA2/WPA3 transition networks would be in play for a long time until WPA3 only access points fully take over.

In initiating the captive portal, the Airgeddon script also initiated a Deauthentication service using Aireplay or MD5 if selected. However, this deauthentication was not successful against the WPA2/WPA3 transition network even though it supported WPA2 configurations. (Dalal et al. 2021)

When the WPA3OpenWrt network was set up to allow only WPA3 SAE only, it showed that connection through WPA2 compatible mobile devices could not be possible resulting in the strength of WPA3. On the cellular device side, the device could not differentiate the encryption either and on that laptop as well. As such the devices label it as unsupported, none or even unspecified encryption.

This research sought to answer the following questions below when incorporated with various ideas to create social engineering attacking platform for WPA3.

- Can WPA3 password be recovered through a captive portal?
- How long can an attack take?
- How effective would this attack be in a real world scenario?

From this research, it can evident to say that the WPA3 password in a WPA2-PSK/WAP3-SAE network can be recovered and that it can be effective in a real world scenario given the capabilities of the devices and software used. The effectiveness of the attack was identified to be dependent on user engagement and the ability of the captive portal to appear legitimate. Due to this, it can be tough to give measurements in terms of time because of the human element. Everyone is different and as such different time frames in determining how effective it is can occur. What can be said though is that people who are not informed of cyber security practices can be enticed or mislead as well as children who may utilize WPA3 networks may not take a second guess of the captive portal and give away the password.

For this experiment to be conducted in a short timeframe,

some assumptions were made. These assumptions are that Protected Management Frame option is not implemented and that the Denial of Service attacks on WPA3 published by various authors such as Mathy Vanhoef, Eyal Ronen, Karim Lounis and Mohammad Zulkernine would work against the WPA3 network since their work has been published and vetted by professional bodies.

One of the assumptions made was that the protected frame management has to be disable in order to capture the handshake via the Dragon Drain attack since the access point CPU cannot handle more than 16 frames/ bad tokens.

As stated before, the use of Bad Tokens can be used to create a Denial of Service on WPA3 as mentioned in the journal Bad Token: Denial of Service attacks on WPA3 by Karim Lounis and

Mohammed Zulkerine, 2019 or the "DragonDrain" exploit derived by Mathy Vanhoef and Eyal Ronen, 2019. These exploits are published and credible modes of deauthentication and are recognized internationally. For this experiment the Dragon drain exploit was chosen to facilitate a deauthentication attack. However due to Time constraints and configuration of these tools, the deauthentication exploit was not created. It must note that the deauthentication attack was only to cause WPA3 devices to be inclined to join the Rouge Access Point when they can't get access to the legitimate Access Point, but it is not necessary to do so.

## V.  Limitations

This research does come with obvious drawbacks. Due to the strength of WPA3 thus far,

Security researchers are working assiduously around the world to enhance and fix all flaws in WPA3. The proposed Denial of Service operates only if certain assumptions are made. These assumptions are: The protected frame management is disable and that access point CPU cannot handle more than 16 frames/ bad tokens.

The Deauthentication attack allows for users to be disconnected so that they can renter the password within the captive portal. However, other approaches to deauthentication still prove useful to achieve the Denial of Service but the attack can be performed without it preying on newly connected devices that don't auto connect to the network.

Cost of devices to be purchased to spawn rouge networks and acquire password is not much and as such devices such as Raspberry Pi's and Wi-Fi pineapple can be configured to run the WPA supplicant but it's still a cost factor.

## VI.  Future Work

Many vulnerabilities to WPA3 has already been disclosed by various security researchers around the globe. These vulnerabilities can work together with other existing exploitation methods to achieve similar desirable results. For example, an attack can utilized Mathy Vanhoef's downgrade attack against a WAP3 system together with a timing side channel attack to acquire information about the WPA3 password. I.e. to determine how complex is the password. Then by using this information, decide whether to use brute force or a captive portal to attack and retrieve the network password.

Future work on WPA3 can be to produce a more resilient Captive Portal to capture WPA3 password. It was noted that the Captive portal was successful when working with a downgrade attack on the WPA2/WPA3 network. However, due to WPA3 devices being able to connect to WPA2 Access Points with ease and not go through the security checks built in for WPA3, Captive Portals might be ideal to accessing WPA2/WPA3 networks by tricking user devices. It was identified that not all devices showed the type of encryption used by the Rouge Access Point. The image below shows a Laptop display of the two WPA3OpenWrt networks side by side. It's only when the icons are scrolled over or clicked for details that the encryption standard is revealed. Not everyone might know of the different types of equipment even though they are using WPA3. This makes the human element vulnerable to social engineering attacks and a Rouge Access Point to be used against them. (Marques et al., 2019)

Also another item that can be looked at in the future is using the strategy of MAC spoofing and the handshake acquired to generate a WPA3 network that acts as an evil twin. Due to time constraints required for this paper, the development of this strategy into a program was affected.

This research will provide insight into the security of next generation Wi-Fi. Once access to WiFi networks is achieved, the risk of penetrating devices on the network increase significantly. Wi-Fi networks are used as a main source of communication amongst devices. It is used from surveillance cameras, operating datacenters, handling Internet of Things (IOT) devices and much more. Gaining access to the Wi-Fi network can also indicate what devices are on the network and scans can be performed to detect which ones are vulnerable to attacks.

## VII.  Conclusion

This research proposes a novel approach to gain the password of a WPA3 network. It is achieved by building on the work produced by many security researchers in order to acquire the WPA3 password in a more effective manner. This research sought to intercept the communication between a client and a WPA2-PSK/WPA3-SAE transition network and acquire the WPA handshake by using a downgrade attack. Then a captive portal is spawned via a rouge WPA2 network with the same SSID where unsuspecting user can enter the password. The entered password hash is then verified using the handshake acquired. Deauthentication attacks were not successful during this experiment that could have resulted from scripts provided not being able to work with the distro utilized. It was also identified that the reverse option of the downgrade attack is possible as well. This is where instead of using a WPA3 access point and a WPA2 device, a WPA3 client device and a WPA2 access point can be used to acquire two fames out of four of the four-way handshake when using a downgrade attack as outlined in the recently discovered flaws of WPA3 by Mathy Vanhoef and Eyal Ronen. Future work in this area can provide significant insight into wireless security against social engineering attacks. Malformed packets were observed to be captured when the HP pavilion laptop tried to connect to the WPA2PSK/WPA3-SAE network. Only laptops that are compatible to connect to the transition network can be utilized even though research articles and the Wi-Fi Alliance claimed that the transition network, WPA2/WPA3, is backwards compatible to work with WPA2. The experiment conducted showed that it is possible to recover the network password via social engineering methods through a captive portal to grant network access..

**Kyle Chadee** is a final year Postgraduate Student at the University of the West Indies, St Augustine campus. He is currently pursuing a Master's Degree in Computer Sciences. He received a bachelor's degree in Computer Science from the University of the West Indies, St Augustine campus, and is interested in wireless networks and computer.

**Wayne Goodridge** received the Ph.D. degree from the Dalhousie University, Halifax, Canada. He is a Senior Lecturer with the Department of Computing and Information Technology, The University of the West Indies, St. Augustine, St. Augustine, Trinidad and Tobago. His research interests include computer communications and security.

**Koffka Khan** received the B.Sc., M.Sc., M.Phil., and D.Phil. degrees from the University of the West Indies. He is currently Lecturer in Computer Science and has up-to-date, published numerous papers in journals & proceedings of international repute. His research areas are Artificial Intelligence, Wireless Computing, Multiobjective Optimization, Stochastic Networks, Evolutionary Computation and Machine Learning.